\documentclass[preprint,superscriptaddress,amsmath,amssymb,aps]{revtex4-1}

\usepackage{graphicx}
\usepackage{dcolumn}
\usepackage{bm}

\begin{document}

\title{Resonant excitation of infra-red emission in GaN:(Mn,Mg)}

\author{Dmytro Kysylychyn} \email{dmytro.kysylychyn@jku.at}
\affiliation{Institut f\"ur Halbleiter-und-Festk\"orperphysik, Johannes Kepler University, Altenbergerstr. 69, A-4040 Linz, Austria}

\author{Jan Suffczy\'nski}
\affiliation{Institute of Experimental Physics, Faculty of Physics, University of Warsaw, Pasteura 5 St., 02-093 Warsaw, Poland}

\author{Tomasz Wo\'zniak}
\affiliation{Department of Theoretical Physics, Faculty of Fundamental Problems of Technology, Wroc\l{}aw University of Science and Technology, PL-50-370 Wroc\l{}aw, Poland}

\author{Nevill Gonzalez Szwacki}
\affiliation{Institute of Theoretical Physics, Faculty of Physics, University of Warsaw, Pasteura 5 St., 02-093 Warsaw, Poland}

\author{Alberta Bonanni} \email{alberta.bonanni@jku.at}
\affiliation{Institut f\"ur Halbleiter-und-Festk\"orperphysik, Johannes Kepler University, Altenbergerstr. 69, A-4040 Linz, Austria}

%\date{\today}

\begin{abstract}
By combining experimental photoluminescence excitation spectroscopy and calculations based on density functional theory and many-body Green's functions, the most efficient excitation channels of infra-red (IR) emission from Mn-Mg$_{k}$ paramagnetic complexes stabilized in GaN:(Mn,Mg) are here identified. Moreover, a Tanabe-Sugano energy diagram for 3$d^{2}$Mn$ ^{5+}$ is reconstructed and Mn-Mg$_{3}$ are singled out as the predominant configurations responsible for the IR emission. The correlation of intensity of the individual emission lines as a function of temperature and excitation energy, allows assigning them to well defined and specific optical transitions.
\end{abstract}

\maketitle

%%%%%%%%%%%%%%%%%%%%
%%% Introduction %%%
%%%%%%%%%%%%%%%%%%%%

\section{\label{sec:intro}Introduction}

Besides their indisputable relevance as building-blocks for high efficiency blue light emitting diodes (LEDs)~\cite{Morkoc:2009_bookV3, Nakamura:2015_RMF,Nakamura:1993_APL, Nakamura:1995_JVSTa} and ultra-violet (UV) laser diodes~\cite{Djavid:2018_PN, Yun:2015_APE, Gutt:2012_APE}, heterostructures of III-nitride semiconductors are currently essential elements in high-electron-mobility transistors~\cite{Del:2017_JMR, Sun:2015_APL}, high-power-~\cite{Hsieh:2016_JEM, Shur:1998_SSE} and spintronic~\cite{Dietl:2015_RMP} devices.

Nitride-based heterostructures active in the infra-red (IR) range of the electro-magnetic spectrum and covering relevant telecommunication wavelengths, are typically alloyed with In~\cite{Mi:2015_PSSb} --challenging the epitaxial fabrication~\cite{Kawakami:2001_JPCM}-- or doped with rare earths, like \textit{e.g.} Er~\cite{Li:2016_OME, Stachowicz:2014_OM, Przybylinska:2001_PB}, significantly affecting the crystallinity of the structures. We have recently demonstrated, that the co-doping --in a metalorganic vapour phase epitaxy (MOVPE) process-- of GaN with Mn and Mg~\cite{Wolos:2004_PRB,Gosk:2005_PRB} induces in the GaN matrix the self-assembly of robust Mn-Mg$ _{k}$ cation complexes~\cite{Devillers:2012_SR, Devillers:2013_APL}, responsible for a broad IR emission~\cite{Devillers:2012_SR, Korotkov:2001_PBCM, Han:2004_APL, Han:2005_APL, Malguth:2007_MRS} around 1~eV (1.2 $\mu$m). Moreover, by growing a codoped GaN:(Mn,Mg) layer hosting Mn-Mg$ _{k}$ complexes on an all-nitride Al$_{x}$Ga$_{1-x}$N:Mn/GaN multilayer half-cavity structure~\cite{Koba:2014_EPL, Devillers:2015_CGD}, the intensity of the IR emission is enhanced significantly~\cite{Capuzzo:2017_SR}. The understanding of the mechanisms underlying the IR emission originating from the Mn-Mg$ _{k}$ complexes in GaN:(Mn,Mg) and the identification of the relevant optical transitions is essential in the perspective of exploiting these material systems in devices with high external quantum efficiency.

By means of photoluminescence excitation spectroscopy, we identify here the most efficient excitation channels of IR emission from Mn$^{5+}$ ions in the Mn-Mg$_{k}$ complexes and we establish unequivocally, that for doping ratios Mg/Mn$\geq$3, Mn-Mg$_{3}$ is the predominant species of Mn-Mg$_{k}$ complexes stabilized in GaN:(Mn,Mg). Moreover, we assign the relevant Mn-Mg$_{3}$-related emission lines to specific and defined transitions contributing to the IR spectrum of GaN:(Mn,Mg).

%%%%%%%%%%%%%%%%%%%%%%%%%%%%%%%%%%%%%%%%%%%%
%%% Experimental and theoretical methods %%%
%%%%%%%%%%%%%%%%%%%%%%%%%%%%%%%%%%%%%%%%%%%%

\section{\label{sec:exp_details}Experimental and theoretical methods}
\paragraph{Samples}
The samples have been grown in an AIXTRON 200RF horizontal tube MOVPE reactor following a protocol previously developed~\cite{Devillers:2015_CGD, Devillers:2013_APL, Capuzzo:2017_SR}. The layers are deposited on $c$-Al$_{2}$O$_{3}$ substrates with TMGa, NH$_{3}$, MnCp$ _{2}$ and Cp$_{2} $Mg as precursors for Ga, N, Mn and Mg, respectively, and employing H$ _{2}$ as carrier gas. Following the nitridation of the sapphire substrate, a GaN low temperature nucleation layer is deposited at 540$^{\circ}$C and then annealed at 975$^{\circ}$C. Subsequently, a 1~$\mu$m-thick wurtzite (wz) GaN buffer is grown at 975$^{\circ}$C, followed by a 600~nm-thick GaN:(Mn,Mg) layer at a substrate temperature of 850$^{\circ}$C. The relative concentration ratio of Mg and Mn is varied along the samples series and reaches $y$=Mg/Mn=4.1, with a total concentration of the dopants $<$1\% cations, as confirmed by secondary ion mass spectroscopy (SIMS). All steps of the epitaxial process are monitored with \textit{in situ} reflectometry and a systematic protocol of structural characterization encompassing atomic force microscopy (AFM), high-resolution x-ray diffraction (HRXRD) and high-resolution transmission electron microscopy (HRTEM), confirms the high crystallinity of the samples. A sketch of the sample structure is provided in the inset to Fig.~\ref{fig:fig1}(a).

\paragraph{Photoluminescence excitation}
For the photoluminescence excitation (PLE) experiments, the samples are kept at $T$~=~7~K in a continuous helium flow cryostat and the measurements are performed in the 1.28~eV--3.06~eV (970~nm--405~nm) energy (wavelength) range. A Mira~900 Ti:sapphire laser working in a continuous wave (cw) mode serves as excitation source between 1.28~eV and 1.77~eV (970~nm--700~nm). An optical parametric oscillator (OPO) allows to extend the laser's range of operation to 1.8~eV--2.25~eV (690~nm--555~nm) and the system is set to a pulsed, picosecond mode. Pulsed and cw excitation with the same average power, are comparably efficient. The excitation between 2.3~eV and 3.06~eV (532~nm--405~nm) is provided either by a set of semiconductor laser diodes --whose outputs are coupled to a conventional optical fiber-- or by an argon-ion laser.

The excitation beam with an integrated power of 7~mW is focused onto the sample surface through a confocal microscope objective (NA = 0.36) to a 1.5~$\mu$m diameter spot, resulting in a power density of $\sim$40~kW/cm$^{-2}$. With increasing excitation energy, the photon flux varies from 1.9$\times$10$^{24}$~s$^{-1}$cm$^{-2}$ to 8.8$\times$10$^{23}$~s$^{-1}$cm$^{-2}$.

A thermo-electrically cooled, single pixel array, (In,Ga)As type CCD camera coupled to a grating monochromator with 1200 grooves/mm is employed as detector. A dichroic (longpass) filter is placed at the entrance of the monochromator, in order to free the signal from stray laser radiation. The overall spectral resolution of the setup attains 0.1 meV.

\paragraph{Theoretical approach}
All calculations are carried out within the density functional theory (DFT) and many-body Green's function Bethe-Salpeter equation (BSE) formalism by using the Vienna $\textit{ab initio}$ simulation package (VASP)~\cite{Hafner:2008_JCC}. The electronic ground state of each Mn-Mg$_{k}$ complex in wz-GaN is calculated using the Perdew-Burke-Ernzerhof (PBE) exchange-correlation functional~\cite{Perdew:1996_PRL} and the projector augmented wave (PAW) method \cite{Bloechl:1994_PRB}. A $\vec{\Gamma}$-centered $\vec{k}$-point mesh of 4$\times$4$\times$4 and a kinetic energy cut-off of 500~eV are employed for Brillouin zone sampling and for expanding the wave functions, respectively. The initial configurations of the Mn-Mg$_{k}$ complexes and the lattice parameters of wz-GaN are taken from Ref.\cite{Devillers:2012_SR}. The ionic positions are optimized with a 0.01~eV/{\AA} tolerance for atomic forces. The absorption coefficient $\alpha\left(\omega\right)$ for each structure is calculated from the dielectric function $\epsilon\left(\omega\right)=\epsilon_{1}\left(\omega\right)+i\epsilon_{2}\left(\omega \right )$ according to:

%%%%%%%%%%%%%%%%%%%%%%%%%%%%%%%%%%%%%%%%%%
\begin{equation}
	\alpha\left(\omega\right)=\omega\sqrt{2\sqrt{\epsilon_1^2\left(\omega\right)+\epsilon_2^2\left(\omega\right)}-2\epsilon_{1}\left(\omega\right)}.
	\label{eq:abs_coeff}
\end{equation}
%%%%%%%%%%%%%%%%%%%%%%%%%%%%%%%%%%%%%%%%%%
 
To compute the imaginary part of the dielectric function, DFT single particle energies moved with a scissors correction of 1.6~eV and the model BSE (mBSE) scheme~\cite{Bokdam:2016_SciRep} are employed. The real part of the dielectric function is then obtained through the Kramers-Kr{\"o}nig relations.

%%%%%%%%%%%%%%%%%%%%%%%%%%%%%%%%%%%%%%%%%%%%%%%%%%
%%% Charge state of Mn in the Mn-Mgk complexes %%%
%%%%%%%%%%%%%%%%%%%%%%%%%%%%%%%%%%%%%%%%%%%%%%%%%%

\section{\label{sec:charge_state} Charge state of M\MakeLowercase{n} in the M\MakeLowercase{n}-M\MakeLowercase{g}$_{k}$ complexes}
\begin{figure}[t]
	\centering
	\includegraphics[width=0.5\textwidth]{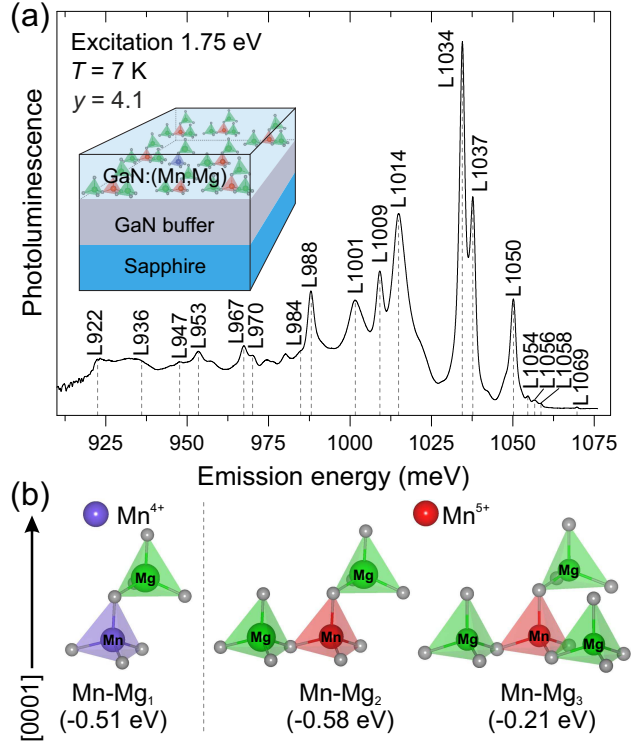}
	\caption{(Color online) (a) IR emission originating from Mn-Mg$_{k}$ complexes in GaN with Mg/Mn ratio $y$=4.1. Selected peaks are labeled according to their emission energy. The sample architecture is sketched in the inset. (b) Most stable Mn-Mg$_{k}$ complexes with $k$=1,2,3 along with their pairing energies. Adapted from Ref.\cite{Devillers:2012_SR}. The charge state of Mn changes from 4+ to 5+, when $k$ is increased from 1 to $k\geq$2~\cite{Devillers:2012_SR,Djermouni:2017_arx}.}
	\label{fig:fig1}
\end{figure}

The photoluminescence (PL) emission from GaN:(Mn,Mg) as a function of energy evidenced in Fig.~\ref{fig:fig1}(a) originates from \textit{d}-shell intra-ion transitions of Mn, whose local environment is influenced by the presence of Mg ions and by the formation of Mn-Mg$_{k}$ complexes. The Mn-Mg$_{k}$ complexes --with the most stable configurations pictured in Fig.~\ref{fig:fig1}(b)-- are deep defects and their energy levels are located in the band-gap of GaN.

As shown in Ref.\cite{Devillers:2012_SR}, the total spin of the Mn cations is modified by the presence of Ga-substitutional Mg ions as nearest neighbors. In particular, the increase of the $y$ relative concentration ratio in the GaN:(Mn,Mg) layer from 0 to 3, changes the total spin of the system gradually from $S$=2 --corresponding to Mn$ ^{3+}$ in GaN:Mn~\cite{Suffczynski:2011_PRB, Bonanni:2011_PRB}-- to $S$=1~\cite{Devillers:2012_SR}, indicating that the majority of the Mn cations changes charge state to Mn$ ^{5+}$. For $y$$<$3, the fraction of Mn-Mg$_{1}$ and Mn-Mg$_{2} $ complexes prevails, while at $y$$\geq$3, Mn-Mg$ _{3}$ complexes become dominant. The dependence of the Mn charge state on the number of Mg cations in the Mn-Mg$ _{k}$ complex has been theoretically confirmed recently~\cite{Djermouni:2017_arx}. The Mn charge state is: (i) 4+ when in the Mn-Mg$ _{1}$ complexes, and (ii) 5+ in the Mn-Mg$_{k}$ with $k>$1, as illustrated in Fig.~\ref{fig:fig1}(b).

Photoluminescence studies as a function of the $y$=Mg/Mn ratio reveal weak IR emission around 1~eV for $y$$<$3, $i.e.$ when Mn-Mg$_{k}$ complexes with $k$=1,2 prevail, while the PL intensity is enhanced by a few orders of magnitude with increasing $y$, reaching a maximum for $y$=4.1, when the majority of the complexes is in the Mn-Mg$ _{3}$ configuration~\cite{Devillers:2012_SR}. Although a similar IR response around 1~eV from GaN:(Mn,Mg) was previously attributed to intra-Mn$^{4+}$ transitions~\cite{Korotkov:2001_PBCM,Han:2004_APL, Han:2005_APL,Malguth:2007_MRS}, the systematic study in Ref.\cite{Devillers:2012_SR} indicates that the mentioned emission is rather related to Mn$^{5+}$ in the Mn-Mg$ _{3}$ complexes. With the present work, we confirm both experimentally and theoretically, the latter assumption and we build an energy diagram for Mn$^{5+}$ in the crystal field of GaN:(Mn,Mg), which enables to assign the single emission peaks to specific intra-ion transitions.

%%%%%%%%%%%%%%%%%%%%%%%%%%%%%%%%%%%%%%%%%%
%%%%%%%%%%%%%%%%%%%%%%%%%%%%%%%%%%%%%%%%%%

%%%%%%%%%%%%%%%%%%%%%%%%%%%%%%%%%%%%
%%% Nature of infra-red emission %%%
%%%%%%%%%%%%%%%%%%%%%%%%%%%%%%%%%%%%

\section{\label{sec:PL} Nature of infra-red emission}
\subsection{\label{subsec:exc_spectra}Resonant excitation}
\begin{figure}[t]
	\centering
	\includegraphics[width=0.5\textwidth]{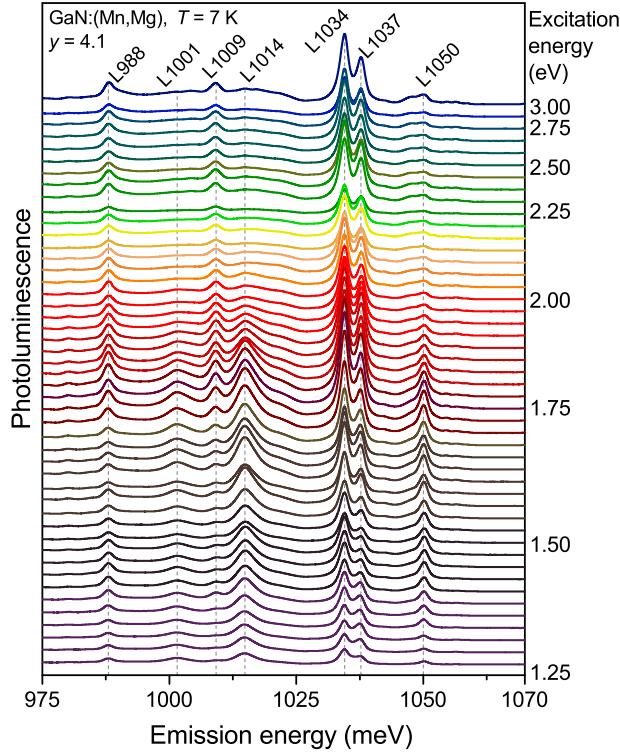}
	\caption{(Color online) Emission spectra of GaN:(Mn,Mg) with $y $=Mg/Mn=4.1 for consecutive excitation energies in a range from 1.28~eV to 3.06~eV at $ T $~=~7~K. Left vertical axis: PL intensity; right vertical axis (non-linear): excitation energy.}
	\label{fig:fig2}
\end{figure}

To shed light on the nature and origin of the transitions in Fig.~\ref{fig:fig1}(a), PLE measurements have been carried out in the range of excitation energies from 1.28~eV to 3.06~eV for GaN:(Mn,Mg) layers grown according to the protocol described in Sec.~\ref{sec:exp_details}, and whose architecture is sketched in the inset to Fig.~\ref{fig:fig1}(a). The emission spectra for a sample with $y$=4.1, --$i.e.$ showing a maximum intensity of the IR emission-- are collected in Fig.~\ref{fig:fig2}, where the excitation energies are expressed in eV, while the emission ones in meV. The spectrum at each excitation energy $E_{\text{exc}}$ is normalized by the number of incident photons $ N_{\text{photons}}$ supplied by the laser beam per time unit: $N_{\text{photons}}=P_{\text{laser}}/E_{\text{exc}}$,
where $P_{\text{laser}}$ is the (constant) laser power density.

%%%%%%%%%%%%%%%%%%%%%%%%%%%%%%%%%%%%%%%%%%
%%%%%%%%%%%%%%%%%%%%%%%%%%%%%%%%%%%%%%%%%%

\begin{figure}[t]
	\centering
	\includegraphics[width=0.5\textwidth]{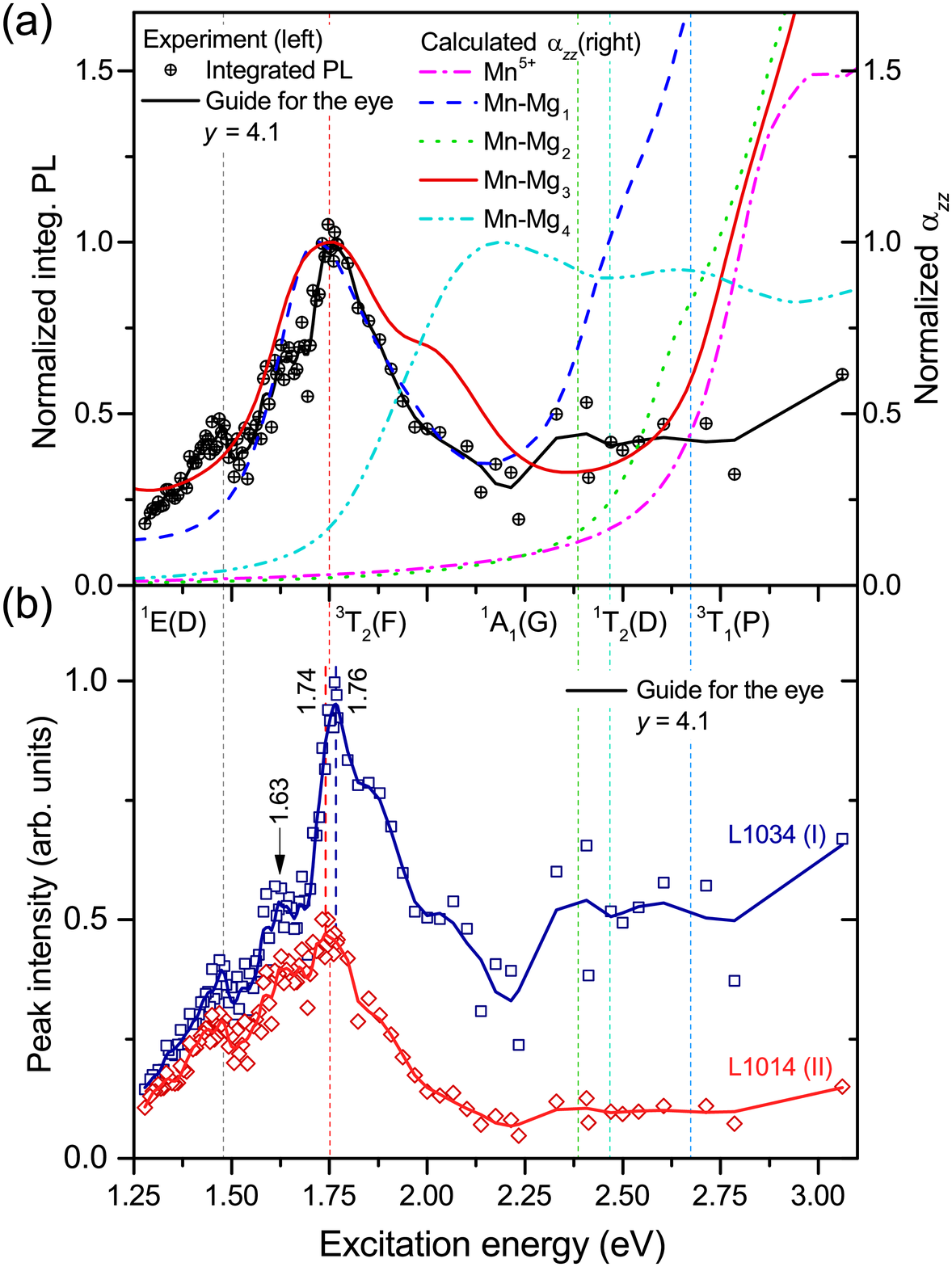}
	\caption{(Color online) (a) Normalized integrated photoluminescence intensity for the considered GaN:(Mn,Mg) active layers with $y$=4.1, as a function of the excitation energy and compared with the calculated absorption coefficient $\alpha_{zz}$ for the Mn$^{5+}$ ion isolated and in the Mn-Mg$ _{1} $, Mn-Mg$ _{2} $, Mn-Mg$ _{3} $, Mn-Mg$ _{4} $ configurations. (b) Integrated PL intensity as a function of the excitation energy for the selected emission lines -representing different behavior groups (I),(II)- centered at 1034.5 and 1014.9~meV, labeled as L1034 and L1014, respectively.} 
	\label{fig:fig3}
\end{figure}

The normalized integrated PL intensity as a function of excitation energy is reported in Fig.~\ref{fig:fig3}(a) (coordinate left axis). The total maximum of the emission is reached for an excitation of 1.75~eV, corresponding to a resonant excitation energy level for the Mn$^{5+}$ ion.

In Fig.~\ref{fig:fig3}(a) (coordinate right axis) also the calculated absorption coefficients $\alpha_{zz}$ for the isolated Mn$^{5+}$ ion and for the various Mn-Mg$_{k}$ configurations ($k$~=~1,2,3,4) in GaN:(Mn,Mg) are reported. The comparison between experimental normalized integrated PL intensity and calculated $\alpha_{zz}$, points at an agreement between the position of the maximum in the experimental data and the calculated one for the Mn-Mg$_{1}$ and Mn-Mg$_{3}$ complexes. In contrast, the $\alpha_{zz}$ for the isolated Mn$^{5+}$-ion without surrounding Mg ions as the nearest neighbors or for Mn$^{5+}$ in the Mn-Mg$_{2}$ complexes, does not present any peaked structure in the considered energy range, ruling out these two configurations as origin of the relevant IR emission. Similarly, the evolution of $\alpha_{zz}$ as a function of energy for the Mn-Mg$ _{4}$ complexes strongly diverges from the experimental trend of the normalized integrated PL intensity, therefore these complexes are also excluded as prevailing sources of the IR luminescence. Finally, although the value of the absorption coefficient of the Mn-Mg$ _{1}$ complexes shows a maximum at $\sim$1.75~eV, a rapid increase of $\alpha_{zz}$ starting at $\sim$2.3~eV does not match the experimental trend, which -- together with the high pairing enthalpy of -0.51~eV for this configuration~\cite{Devillers:2012_SR}-- excludes it from being a likely origin of the IR transitions. These considerations allow to identify the Mn-Mg$ _{3}$ complexes as responsible for the observed IR emission. The deviation of the theoretical curves from the experimental ones results from limitations in the calculation accuracy --due to the reduced dimensions of the cell computed-- and to uncertainty in the efficiency of the relaxation mechanisms from the absorbing to the emitting state.

Depending on the response to the variations in excitation energy of the emission peaks labeled in Fig.~\ref{fig:fig1}(a), the following trends are identified: (I) peaks L988, L1009, L1034 and L1037 persist at all considered excitation energies; (II) peaks L1001, L1014 and L1050 significantly quench at excitation energies above 2~eV. This suggests, that the emitting levels within a group share a common specific excited state.

As evidenced in Fig.~\ref{fig:fig3}(b), the peak intensities of L1034 and L1014: (i) reach their resonant excitation at 1.76~eV and 1.74~eV, respectively, and (ii) show a different behavior as a function of the excitation energy, which is assigned to the diverse local environment of the two emitting Mn-centers. Moreover, at excitation energies above the resonant one, the intensity of L1034 and L1014 is quenched by a factor of two for L1034 and, significantly, by a factor five for L1014, confirming the specific response of the two emitting centers, sharing the same charge state, but differing in the local environment or arrangement of the Mg ions in the complex.

%%%%%%%%%%%%%%%%%%%%%%%%%%%%%%%%%%%%%%%%%%
%%%%%%%%%%%%%%%%%%%%%%%%%%%%%%%%%%%%%%%%%%

\subsection{\label{subsec:en_diagram}Energy diagram of Mn$ ^{5+} $ in GaN:(Mn,Mg)}
\begin{figure}[t]
	\centering
	\includegraphics[width=0.5\textwidth]{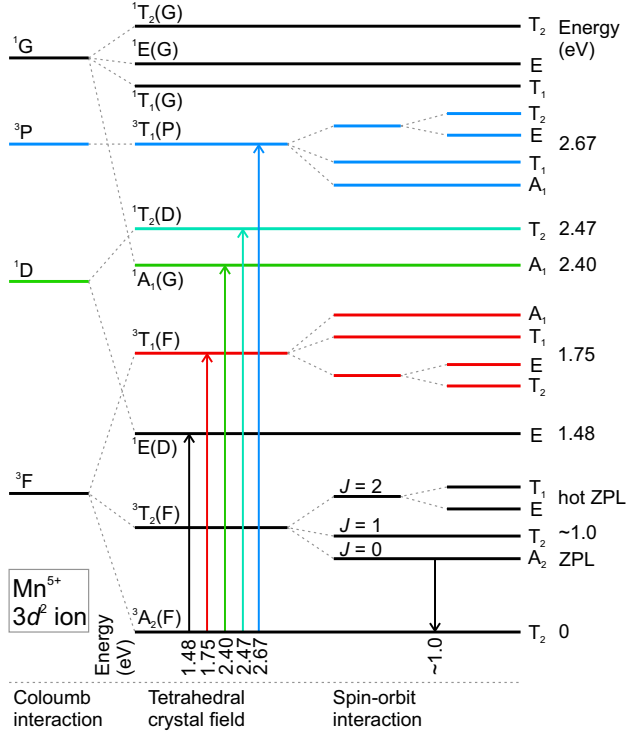}
	\caption{(Color online) Energy diagram for 3$d^{2}$Mn$ ^{5+}$ substitutional of Ga in the crystal field of GaN reconstructed by employing: (i) a Tanabe-Sugano diagram for $d^{2}$-ions in a tetrahedral configuration, (ii) excitation and (iii) emission spectra of Mn$ ^{5+} $ in a Mn-Mg$ _{3}$ complex.  The strength of the calculated crystal field is $Dq  $ = 833 cm$ ^{-1}$. Splitting due to spin-orbit interaction is accounted for, neglecting the one due to the breaking of the trigonal crystal symmetry.}
	\label{fig:fig4}
\end{figure}

The energy diagram for the 3$d ^{2}$ configuration of specific ions in GaN was already reported for Ti$^{2+}$~\cite{Heitz:1995_PRB}, V$^{3+}$~\cite{Baur:1995_APL,Thurian:1997_APL, Kuck:1999_OM}, and Cr$^{4+}$~\cite{Baur:1995_APL,Shen:2000_PRL}. Here, for the particular case of 3$d^{2}$ Mn$^{5+}$, we follow the Tanabe-Sugano approach --based on the occupancy of the $d$-shell of the transition metal and on symmetry considerations-- accounting for the splitting of the energy levels, due to the crystal field ~\cite{Sugano:1970_PNY}. Specifically, we consider Mn in GaN in the tetrahedrally coordinated Mn-N$ _{4}$ arrangement, and we reconstruct the energy diagram reported in Fig.~\ref{fig:fig4}, based on the Tanabe-Sugano scheme and on the obtained experimental excitation and emission spectra.

In the case of the studied structures, the strength of the tetrahedral crystal field splitting is estimated to be $Dq$=833~cm$^{-1}$, as obtained graphically by means of the Tanabe-Sugano diagram. The reliability of the method is supported by reference calculations of the crystal field $Dq$ for V$ ^{3+} $ and Cr$ ^{4+} $ in GaN and confirmed with the values in Ref.~\cite{Baur:1995_APL}. Due to spin-orbit coupling, the state $^{3}$T$_{2}$(F) with energy $E$ splits into: (1) a singlet ($J$=0), a triplet ($J$=1) and a quintet ($J$=2). The latter splits further in the cubic crystal field, due to  off-diagonal elements of the spin-orbit interaction with other excited terms, into a  doublet E and a triplet  T. Moreover, for the Mn-N$_{4} $ arrangement in the wz structure the contribution of the trigonal crystal field --with magnitude comparable to the one of the spin-orbit coupling-- has to be taken into account. Additionally, in a Mn-Mg$ _{k}$ complex, due to the shortening of the Mn-N bonds caused by the presence of Mg~\cite{Devillers:2012_SR}, the symmetry is lower than trigonal, except for the Mn-Mg$ _{3}$ case, which arranges all the relevant Mg ions in the (0001) plane. This implies that complexes in different configurations exhibit different symmetries, hence, resulting in energy diagrams that are relatively shifted.

%%%%%%%%%%%%%%%%%%%%%%%%%%%%%%%%%%%%%%%%%%
%%%%%%%%%%%%%%%%%%%%%%%%%%%%%%%%%%%%%%%%%%

The strongest resonant excitation occurring at 1.75~eV in Fig.~\ref{fig:fig3}(a) is ascribed to the absorption of incoming photons with energy corresponding to the $^{3}$A$_{2}$(F)$\rightarrow$$^{3}$T$_{1}$(F) transition illustrated in Fig.~\ref{fig:fig4}. A subsequent relaxation occurs to the lowest excited state $^{3}$T$_{2}$(F), followed by a radiative transition to the ground state $^{3}$A$_{2}$(F) with the emission of an IR photon with energy $E$$\sim$1000~meV. This shows that the excitation is the most efficient when both, absorbing and emitting level, share the same symmetry and indicates that the relaxation process preserves the symmetry of the absorbing state.

Other resonant excitations -reported in Fig.~\ref{fig:fig3}(a) and Fig.~\ref{fig:fig4}- take place due to transitions from the ground state $^{3}$A$_{2}$(F): (i) to the $^{1}$E(D) level with energy $\sim$1.48 eV, and (ii) to $^{1}$A$_{1}$(G), $^{1}$T$_{2}$(D), $^{3}$T$_{1}$(P) states with specific energies that are predicted by the Tanabe-Sugano diagram: $\sim$2.40~eV, $\sim$2.47~eV, $\sim$2.67~eV, respectively. Since these levels exhibit a different symmetry than the emitting $^{3}$T$_{2}$(F) state, the relaxation from them to the $^{3}$T$_{2}$(F) level and the resulting IR emission are less efficient than in the case of excitations through $^{3}$T$_{1}$(F) level. The feature at $\sim$1.63~eV in the PLE spectrum of Fig.~\ref{fig:fig3}(b) is assigned to a total splitting of $^{3}$T$_{1}$(F) with contributions from: (i) spin-orbit interaction introducing a splitting of the order of 40-60~eV~\cite{Baur:1995_APL}, and (ii) splitting from the breaking of the trigonal symmetry (not shown in Fig.~\ref{fig:fig4}).

\subsection{\label{subsec:peaks_origin}Assignment of the emission lines}

\begin{figure}[t]
	\centering
	\includegraphics[width=0.4\textwidth]{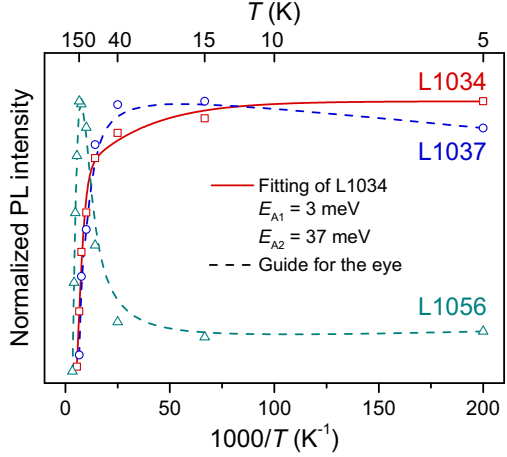}
	\caption{(Color online) Normalized PL intensity for L1034, L1037 and L1056 as a function of inverse temperature; top axis: temperature. Up to 40~K the intensity of L1037 increases, due to thermally activated hopping of electrons from the L1034 level to the L1037 one. The enhancement of intensity with temperature points at L1056 as hot-ZPL.}
	\label{fig:fig5}
\end{figure}

The two most intense emission peaks reported in Fig.~\ref{fig:fig1}(a), namely L1034 and L1037, show a comparable behavior as a function of the excitation energy and a commensurate value of full-width-at-half-maximum (FWHM) of $\sim$2~meV. These peaks are not due to transitions from the same excited level to the $^{3}$A$_{2}$(F) ground state split by the trigonal distortion of the complex, since the strength of this splitting --being 0.31 meV-- is one order of magnitude smaller than their relative energy difference 3~meV, as reported in Ref.\cite{Devillers:2012_SR}. The temperature dependence of the emission reveals an ebbing of L1034 and an enhancement of the emission intensity of L1037 up to 40~K --as reported in in Fig.~\ref{fig:fig5}-- resulting from a Boltzmann redistribution of the population of levels. In contrast, above 40~K non-radiative processes quench the emission. This behavior implies that the initial state of the L1034 and L1037 transitions: (i) belongs to the same Mn-center, and (ii) is separated by 3~meV, due to the splitting originating from the distortion of the Mn-Mg$ _{3}$ complex in the trigonal field. Moreover, it indicates that the lifetime of the excited state is much longer than the relaxation between the levels. Thus, the attenuation of the L1034 intensity can be described assuming the presence of two competing non-radiative processes~\cite{Leroux:1999_JAP} according to the relation for the intensity $I$:

%%%%%%%%%%%%%%%%%%%%%%%%%%%%%%%%%%%%%%%%%%
\begin{equation}
	I(T)=\frac{I_0}{1+A_1\exp(-\frac{E_\mathrm{A1}}{k_\mathrm{B}T})+A_2\exp(-\frac{E_\mathrm{A2}}{k_\mathrm{B}T})},
	\label{eq:therm_pop}
\end{equation}
%%%%%%%%%%%%%%%%%%%%%%%%%%%%%%%%%%%%%%%%%%

with $k_\mathrm{B}$ being the Boltzmann constant, $T$ the temperature and with the activation energies $E_\mathrm{A1}$ and $E_\mathrm{A2}$ of the two competing processes calculated to be 3~meV and 37~meV, respectively. The estimated activation energy $E_\mathrm{A1}$=3~meV corresponds to the energy separation of the two levels from which the emission lines L1034 and L1037 arise. The activation energy $E_\mathrm{A2}$=37~meV matches the magnitude of the splittings due to spin-orbit interaction and to the trigonal symmetry distortion.

We identify L1054, L1056, L1058 and L1069 reported in Fig.~\ref{fig:fig1}(a) as hot zero phonon lines (hot-ZPLs), since their intensity significantly increases with temperature as reported in Fig.~\ref{fig:fig5} for L1056, pointing at transitions from upper levels of the splitted $ ^{3}$T$_{2}$(F) state for diverse orientations of the Mn-Mg$ _{3} $ complexes. In contrast, the intensity of the remaining emission lines is quenched with increasing temperature.

%%%%%%%%%%%%%%%%%%%%%%%%%%%%%%%%%%%%%%%%%%
%%%%%%%%%%%%%%%%%%%%%%%%%%%%%%%%%%%%%%%%%%

The transitions L988, L1001, L1009, L1014, L1034 and L1050, are attributed to ZPLs corresponding to a well defined Mn-Mg$ _{3}$ complex with diverse orientations, resulting in a shift of the energy levels due to the trigonal distortion specific to each of them.

\begin{figure}[t]
	\centering
	\includegraphics[width=0.5\textwidth]{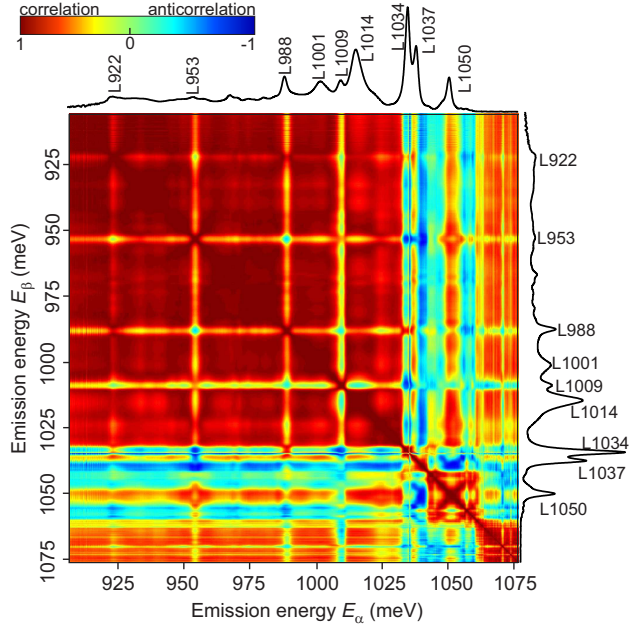}
	\caption{(Color online) Map of the $ \varGamma $ correlation coefficient matrix for spectra measured at different excitation energy. The color scale identifies the correlation $ \varGamma > 0$  (red) and anticorrelation $ \varGamma < 0 $ (blue) of the emission lines at energies $E_{\alpha}$ and $E_{\beta}$, respectively. The average of all recorded PL spectra is shown along the top horizontal axis and along the right-hand side vertical axis.}
	\label{fig:fig6}
\end{figure}

In order to confirm the identification of the emission lines related to the same orientation of the complex and the assignment of phonon replicas, the correlation factor $\varGamma$ is evaluated, according to~\cite{Pietka:2013_PRB}:

%%%%%%%%%%%%%%%%%%%%%%%%%%%%%%%%%%%%%%%%%%
\begin{equation}
	\varGamma = \frac{\sum_{i} (I_{i}^{\alpha}-\bar{I}^{\alpha}) (I_{i}^{\beta}-\bar{I}^{\beta})}{\sqrt{\sum_{i} (I_{i}^{\alpha}-\bar{I}^{\alpha})^{2} \sum_{i}(I_{i}^{\beta}-\bar{I}^{\beta})^{2}}},
	\label{eq:correlation}
\end{equation}
%%%%%%%%%%%%%%%%%%%%%%%%%%%%%%%%%%%%%%%%%%

where $I_{i}^{\alpha}$ and $ I_{i}^{\beta} $ indicate the intensity of the signal in the spectrum $i$ at the emission energy $E_{\alpha}$ and $E_{\beta}$, respectively; $\bar{I}^{\alpha}$ and $\bar{I}^{\beta}$ - the intensity averaged over all spectra at the emission energy $E_{\alpha}$ and $E_{\beta}$, respectively. The index $i$ varies from 1 to 101 and labels the emission spectra recorded for subsequent excitation energies. The correlation map for the emission in the energy range between 906~meV and 1076~meV is reproduced in Fig.~\ref{fig:fig6}. When emissions at two selected energies $E_{\alpha}$ and $E_{\beta}$ are correlated, $i.e.$ $\varGamma \rightarrow 1$ (anti-correlated $ \varGamma \rightarrow -1$) they show the same (inverse) behavior as a function of the excitation energy: if the intensity of the $E_{\alpha}$-line is enhanced, the one of the $E_{\beta}$-line increases as well (decreases). If $\varGamma\rightarrow$~0, the two emissions are not correlated.

%%%%%%%%%%%%%%%%%%%%%%%%%%%%%%%%%%%%%%%%%%
%%%%%%%%%%%%%%%%%%%%%%%%%%%%%%%%%%%%%%%%%%

For our case, the correlated emission lines --$i.e.$ showing similar response to variations of temperature and excitation energy-- are collected and detailed in Table~\ref{table:corr}. The significant emission lines between 988~meV and 1050~meV, correlate with their phonon replica $A_{1}$(TO), separated by 66~meV. Furthermore, peak L988 correlates with L1034, though their separation in energy of 46.5~meV cannot be traced back to any phonons. A similar correlation is observed for the pairs L953--L1009, and L1001--L1050. Their coherent response points to transitions between the energy levels with equal symmetries in similarly oriented Mn-Mg$ _{3}$ complexes, but with different Mn-environment, \textit{i.e.} with different bond lengths and resulting in a relative shift of the energy levels. The emissions L1034 and L1014 show no reciprocal correlation and a different behavior as a function of the excitation energy -- as evidenced in Fig.~\ref{fig:fig3}(b) -- indicates that these lines are likely to originate from differently oriented complexes that have specific absorption cross-sections and diverse radiation diagrams.

%%%%%%%%%%%%%%%%%%%%%%%%%%%%%%%%%%%%%%%%%%
\begin{table*}
\centering	
	\begin{tabular*}{\textwidth}{l@{\extracolsep{\fill}}cccc}
		\hline \hline
		Line & Energy (meV) & Correlation with & $ \Delta E $ (meV) & Assignment \\ \hline
		L922 & 922.5& L988& 65.5& $A_{1}$(TO) phonon replica\\
		L936 & 936.0& L1001& 65.5& $A_{1}$(TO) phonon replica\\
		L947 & 947.7& L1014& 67.2& $A_{1}$(TO) phonon replica\\
		L953 & 953.5& L1009& 55.7& -- \\
		L967 & 967.4& L1034& 67.1& $A_{1}$(TO) phonon replica\\
		L970 & 970.1& L1037& 67.6& $A_{1}$(TO) phonon replica\\
		L984 & 984.8& L1050& 65.4& $A_{1}$(TO) phonon replica\\
		L988 & 980.0& L1034& 46.5& different configuration, ZPL \\
		L1001 & 1001.5& L1050 & 48.7& different configuration, ZPL \\
		L1009 & 1009.2& -- & -- & ZPL \\
		L1014 & 1014.9& L1050 & 35.5 & different configuration, ZPL \\
		L1034 & 1034.5 & -- & -- & ZPL \\
		L1050 & 1050.2 & -- & -- & ZPL \\
		\hline \hline
	\end{tabular*}
\caption{Correlation between various PL emission peaks and their assignment. A different configuration implies a different orientation of the Mn-Mg$ _{3} $ complexes producing a specific emission line.}
\label{table:corr}
\end{table*}
%%%%%%%%%%%%%%%%%%%%%%%%%%%%%%%%%%%%%%%%%%

%%%%%%%%%%%%%%%%%%%%%%%%%%%%%%
%%% Discussion and outlook %%%
%%%%%%%%%%%%%%%%%%%%%%%%%%%%%%

\section{\label{sec:discussion}Discussion and outlook}
The IR emission from GaN:(Mn,Mg) epitaxial layers hosting Mn-Mg$ _{k}$ complexes has been investigated $via$ photoluminescence excitation spectroscopy. Intra-ion transitions from Mn$^{5+}$ in the Mn-Mg$ _{k}$ complexes are found to play a dominant role in the origin of the IR emission. The resonant PLE observed at 1.75~eV is identified as the transition from the $^{3}$A$_{2}$(F) ground state to the $^{3}$T$_{1}$(F) one in the energy diagram reconstructed through a Tanabe-Sugano diagram for 3$d^{2}$ tetrahedrally coordinated ions. The calculated absorption coefficient $\alpha_{zz}$ for the relevant Mn-Mg$ _{k}$ complexes with $k$=1-4 reveals a maximum in the absorption for Mn-Mg$_{3}$ complexes at the resonant excitation of the IR emission. This supports the premise of Mn-Mg$_{3}$ complexes --stabilized in various orientations with respect to the host crystal-- being responsible for the observed IR emission. Furthermore, trends in the relative intensity changes of selected emission lines with the excitation energy, are identified and assigned to the different radiation diagrams of the related Mn-Mg$ _{3} $ complexes. Moreover, the correlation of the most intense emission lines with the corresponding $A_{1}$(TO) phonon replicas has been established.

Besides opening perspectives for the realization of alternative nitride-based IR (single) emitters through the understanding of the emission mechanisms in correlation with the growth protocols, this study contributes to build a platform for the analysis of coherent phenomena involving the spin degree of freedom of the paramagnetic complexes. The understanding of the excitation and emission mechanisms in the considered complexes, is also significant in the view of implementing these stable self-assembled sub-nano-objects in $e.g.$ spin memories, where information may be written in the spin state of the ion by polarized optical excitation, stored and then read-out either optically~\cite{Smolenski:2016_NatCommun} or electrically~\cite{Dietl:2014_RMP}.

%%%%%%%%%%%%%%%%%%%%%%%
%%% Acknowledgments %%%
%%%%%%%%%%%%%%%%%%%%%%%

\section*{\label{sec:acknowledgments}Acknowledgments}
This work was supported by the EC Horizon2020 Research and Innovation Programme (grant 645776), by the Austrian Science Foundation - FWF (P24471 and P26830), and by the Austrian Agency for International Cooperation in Education and Research (PL 01/2017). Special thanks to Hanka Przybylinska for fruitful discussions.
TW and NGSz gratefully acknowledge the support of the National Science Centre (NCN) through the grant UMO-2013/11/B/ST3/04273. Numerical calculations were performed at ICM at the University of Warsaw.

%%%%%%%%%%%%%%%%%%%%
%%% Bibliography %%%
%%%%%%%%%%%%%%%%%%%%

%\bibliography{Kysylychyn_bibliography_PLE}
\providecommand{\noopsort}[1]{}\providecommand{\singleletter}[1]{#1}%

\end{document}